\documentclass[preprint, showpacs,preprintnumbers,amsmath,amssymb]{revtex4}
\usepackage[dvips]{graphicx}
\usepackage{graphicx}
\usepackage{amsfonts}
\usepackage{bm}
\usepackage{amsmath}
\usepackage{amssymb}
\usepackage{color}
\usepackage[all]{xy}

\def\be{\begin{equation}}
\def\ee{\end{equation}}
\def\bea{\begin{eqnarray}}
\def\eea{\end{eqnarray}}

\begin{document}

\title{Geometrothermodynamics and critical behavior of regular black holes}

\author{Hernando Quevedo}
\email{quevedo@nucleares.unam.mx}
\affiliation{Instituto de Ciencias Nucleares, Universidad Nacional Aut\'onoma de M\'exico, Mexico}
\affiliation{Dipartimento di Fisica and Icra, Universit\`a di Roma “La Sapienza”, Roma, Italy}
\affiliation{Al-Farabi Kazakh National University, Almaty, Kazakhstan}

\author{Mar\'ia N. Quevedo}
\email{maria.quevedo@unimilitar.edu.co} \affiliation{Departamento
de Matem\'aticas, Facultad de Ciencias B\'asicas,\\Universidad
Militar Nueva Granada, Cra 11 No. 101-80, Bogot\'a D.C., Colombia}

\author{Alberto S\'anchez}
\email{asanchez@ciidet.edu.mx} \affiliation{Departamento de
Posgrado, CIIDET,\\{\it AP752}, Quer\'etaro, QRO 76000, MEXICO}

\date{\today}

\begin{abstract}

 We apply the formalism of quasi-homogeneous geometrothermodynamics to study the stability, phase transition, and criticality properties of the Bardeen regular black hole. We show that the singularities of thermodynamic curvature determine the regions where the stability conditions break down and indicate the location of the Davies phase transition and the zero-temperature limiting case. Moreover, we compute the critical exponents of the curvature and prove the existence of a scaling relation with the heat capacity.

{\bf Keywords:} Thermodynamics, geometrothermodynamics, phase transitions, regular black holes

\end{abstract}

\pacs{05.70.Ce; 05.70.Fh; 04.70.-s; 04.20.-q}

\maketitle

\section{Introduction}
\label{intro}

Regular black holes are defined as exact solutions of gravity theories for which the curvature is free of singularities in the entire spacetime. In 1968, Bardeen \cite{bardeen} obtained a regular solution that can be interpreted as a gravitationally collapsed magnetic monopole
arising in a specific form of nonlinear
electrodynamics \cite{eloy}.
Later on, 
Hayward in
\cite{hayward} and Dymnikova in \cite{dymnikova} derived alternative black hole solutions without curvature singularities.  Hayward solution is interpreted as describing the formation of a black hole from an
initial vacuum region with finite density and pressure, which vanishes rapidly at large distances and behaves as a cosmological
constant at small distances. Dymnikova black hole is a
non-singular spherically symmetric solution, which has been also generalized to include a non-singular cosmological black hole solution \cite{dymnikova2}.

Recently, regular black hole solutions have been intensively studied as possible candidates to describe the gravitational field of astrophysical compact objects \cite{Bambi,Ghosh,Toshmatov,Ghosh2,Neves}. 
For this reason, these solutions have been used
to study singularity problems, quasi-normal modes of black
holes \cite{Lin}, geodesic motion \cite{Chiba},
wormholes \cite{Halilsoy,Kuhfittig,Sharif}, black hole
thermodynamics \cite{Maluf}, strong deflection lensing \cite{Zhao}, and several effects in Einstein-Gauss-Bonnet gravity \cite{Arun}.

On the other hand, since the works of Davies \cite{davies}, Smarr
\cite{smarr}, and Bekenstein \cite{beke}, black hole
thermodynamics has been the subject of numerous researches in
theoretical physics, due to its possible connection to a 
still-unknown theory of
quantum gravity, which could allow us to understand the microscopic structure of black holes. Moreover, an alternative approach to black hole
thermodynamics consists in using differential geometry to study
the properties of the particular equilibrium manifold, whose points represent equilibrium states of the corresponding system. In this way, the geometric properties of the equilibrium space turn out to be related to the thermodynamic properties of the system.

The study of black hole thermodynamics and its
relationship with geometry has been the subject of intensive
research \cite{Amari,Bravetti,Aman,Aman2,Aman3,shen,Cai}. This geometric
study has been considered in several works by using  different
approaches \cite{Weinhold,Ruppeiner,quevedo2}. The most recent of
these approaches is called geometrothermodynamics (GTD)
\cite{quevedo2}, which is
is  a formalism that relates the 
contact structure of  the phase space $\mathcal{T}$ with the
metric structure of the equilibrium space $\mathcal{E}$, which is a special subspace of $\mathcal{T}$. One of the essential features of GTD is that it incorporates into the geometric approach the concept of Legendre invariance, which is a central aspect in ordinary thermodynamics and represents the fact that the properties of a system do not depend on the thermodynamic potential used for its description.

In this work, we will
use  Legendre invariant metrics in the context of GTD to construct an invariant geometric
representation of the thermodynamics of regular black holes. We will focus on the Bardeen spacetime as a representative of the family of regular black hole solutions.

This paper is organized as follows. In Sec. \ref{sec:revgtd},  we review the
fundamentals of GTD, emphasizing the explicit expressions and the role of the Legendre invariant metrics. In Sec. \ref{sec:bardeen}, we study the most important
aspects of the Bardeen regular black hole, emphasizing the thermodynamic
interpretation of its physical parameters.  In Sec. \ref{sec:gtd}, we
apply the formalism of GTD to the case of the Bardeen regular black hole and show that the GTD metrics contain all the information about the stability properties and phase transitions of the black hole. In Sec. \ref{sec:inter}, we interpret our results and show the compatibility between the findings obtained in GTD and those of ordinary thermodynamics. 
Section \ref{sec:cri} is dedicated to the analysis of the  behavior of the Bardeen black hole close to the critical point, where a phase transition takes place.
Finally, in Sec. \ref{sec:con}, we present the
conclusions of our work.


\section{Geometrothermodynamics} \label{sec:revgtd}

GTD is a formalism that represents the properties of thermodynamic systems in terms of geometric concepts, taking into account the fact that ordinary thermodynamics is invariant with respect to Legendre transformations 
\cite{quevedo2}. In physical terms, this invariance means that the properties of a thermodynamic system do not depend on the choice of thermodynamic potential used for its description \cite{callen}. To this end, GTD represents Legendre transformations as coordinate transformations, which are defined on a $(2n+1)$--dimensional manifold $\mathcal{T}$ with a set of
coordinates $Z^A =\{\Phi, E^a, I_a\}$, where $\Phi$ represents any thermodynamic potential and $E^a$ ($I_a$), $n=1,2,...,n$ are related to the extensive (intensive) variables that are needed to describe the system. However, the coordinates of  ${\cal T}$ are independent of each other in order to be able to define Legendre transformations in terms of the coordinates of ${\cal T}$ as 
\cite{arnold,robert} 
\bea
\label{trans1} &&\Big\{Z^A\Big\}\longrightarrow
\Big\{\tilde{Z}^A\Big\}=\Big\{\tilde{\Phi}\,,\tilde{E}^a\,,\tilde{I}^a\,,\Big\}\\\label{trans2}
\Phi &=&\tilde{\Phi}-\tilde{E}^k\tilde{I}_k\,,\quad
E^i=-\tilde{I}^i\,,\quad E^j=\tilde{E}^j\,,\quad
I^j=\tilde{I}^j\,,\quad E^i=-\tilde{I}^i\,,
\eea 
where $i\cup j$ is
any disjoint decomposition of the set of indices ${1,...,n}$, and
$k,l = 1,...,i$ so that for $i = {1,...,n}$ and $i =
\varnothing$, we obtain the total Legendre transformation and the
identity, respectively. We assume that ${\cal T}$ is a differential manifold; consequently, we can endow it with a metric $G_{AB}$, which can depend explicitly on the coordinates $Z^A$. The Legendre invariance of ${\cal T}$ is guaranteed if all the geometric objects defined on it are invariant with respect to the coordinate transformations (\ref{trans1}) and (\ref{trans2}). In particular, we demand that the functional dependence of the components $G_{AB}$ remain unchanged under the action of Legendre transformations. 
It turns out that this condition is satisfied by the following line elements
\cite{quasihomo,vander}
\be
G^{^{I}}=  (d\Phi - I_a d E^a)^2 + (\xi_{ab} E^a I^b) (\delta_{cd} dE^c dI^d) \ ,
\label{GI}
\ee
\be 
G^{^{II}}= (d\Phi - I_a d E^a)^2 + (\xi_{ab} E^a I^b) (\eta_{cd} dE^c dI^d) \ ,
\label{GII}
\ee
\be	
\label{GIII}
G^{{III}}  =(d\Phi - I_a d E^a)^2  + \sum_{a=1}^n \xi_a (E^a I^a)^{2k+1}   d E^a   d I^a \ ,
\ee
where $\eta_{ab}= {\rm diag}(-1,1,\cdots,1)$, $\xi_a$ are real constants, $\xi_{ab}$ is a diagonal $n\times n$ real matrix, and $k$ is an integer. 

In addition, the odd-dimensional differential manifold ${\cal T}$ allows the introduction of the canonical contact 1-form $\Theta = d\Phi - I_a d E^a$, which is also Legendre invariant, i.e., under the action of a Legendre transformation $Z^A\rightarrow \tilde Z ^ A$, it does not change its functional dependence: $\Theta \rightarrow \tilde \Theta = d \tilde \Phi - \tilde I_ a \tilde E^a $.  This ends the construction of the phase space of GTD, which is defined as the Legendre invariant triad $({\cal T},\Theta, G) $. 

The second ingredient of GTD is the equilibrium space $\mathcal{E}$, which is defined as a subspace of $\mathcal{T}$ by means of a smooth embedding map $\varphi: \mathcal{E}\to \mathcal{T}$ such that $\varphi^*(\Theta)=0$, where $\varphi^*$ is the corresponding pullback. In terms of coordinates, the embedding map $\varphi$ implies that $Z^A = \{ \Phi(E^a), E^a, I_a(E^b)\}$. Then, the condition 
$\varphi^*(\Theta)=0$ implies that on the equilibrium space $d\Phi = I_a dE^a$ with  $I_a = \frac{\partial \Phi}{\partial E^a} $, which corresponds to the first law of thermodynamics in the GTD representation.
The line element $G= G_{AB} dZ^ A d Z^B$ on $\mathcal{T}$ induces on ${\cal E}$ a line element $g=g_{ab}dE^a dE^b$ by means of the pullback, i.e., $\varphi^*(G)=g$.
Then, from Eqs.(\ref{GI}), (\ref{GII}), and (\ref{GIII}), we obtain \cite{quasihomo}
\be
g^{{I}} =  \sum_{a,b,c=1}^n \left( \beta_c E^c \frac{\partial\Phi}{\partial E^c} \right)  
\frac{\partial^2\Phi}{\partial E^a \partial E^b}  dE^ a d E^ b ,
\label{gdownI}
\ee
\be
g^{II} =   
 \sum_{a,b,c,d=1}^n \left( \beta_c E^c \frac{\partial\Phi}{\partial E^c} \right) 
 \eta_a^{\ d}
\frac{\partial^2\Phi}{\partial E^b \partial E^d} dE^ a d E^ b   ,
\label{gdownII}
\ee
\be
g^{{III}} = \sum_{a,b=1} ^n \left( \beta_a  E^a \frac{\partial\Phi}{\partial E^a}\right)
 \frac{\partial ^2 \Phi}{\partial E^a \partial E^b}
dE^a dE^b \ ,
\label{gdownIII}
\ee
respectively, 
where $\eta_a^{\ c}={\rm diag}(-1,1,\cdots,1)$. The free parameters of the line elements $G^I$, $G^{II}$, and $G^{III}$ have 
been chosen as $\xi_a=\beta_a$ and $\xi_{ab} = {\rm diag}(\beta_1,\cdots,\beta_n)$, where the parameters $\beta_a$ are the quasi-homogeneous coefficients of the fundamental equation $\Phi=\Phi(E^a)$, i.e., the constants $\beta_a$ that satisfy the condition $\Phi(\lambda^{\beta_a}E^a)=\beta_\Phi \Phi(E^a)$, where $\lambda$ is a real constant. The constant $k$ can be chosen as $k=0$ by demanding that the metrics $g^I_{ab}$, $g^{II}_{ab}$, and $g^{III}_{ab}$, can be applied to the same thermodynamic system simultaneously, leading to compatible results. Moreover, the Euler identity, $\sum \beta_a E^a I_a = \beta_\Phi \Phi$, can be used to further simplify the expressions of the GTD metrics.  

The GTD approach consists in calculating the components of the metrics of the equilibrium space for a specific thermodynamic system \cite{vander}, analyzing the corresponding geometric properties, in particular their curvature singularities in order to determine the correspondence with the phase transition structure of the system. 


\section{Bardeen regular  black hole and its thermodynamics}
\label{sec:bardeen}

The spherically symmetric Bardeen black hole is described by  the
metric \cite{bardeen,orlando}

\bea \label{metric} ds^2=-f(r)dt^2+f(r)^{-1}dr^2+r^2(d\theta^2
\sin^2 \theta d\varphi^2)\,,\eea where
\be \label{lapse}
f(r)=1-\frac{2Mr^2}{(r^2+q^2)^{\frac{3}{2}}}\,,\ee
with $q$ and
$M$ representing the magnetic charge and the mass of the black hole,
respectively.

The roots of the lapse function ($g_{tt} = f = 0$) define the horizons $r = r_\pm$ of the spacetime. Moreover, the null hypersurface $r = r_+$
corresponds to an event horizon, which in this case is also a Killing horizon,
and the inner horizon at $ r_-$ is a Cauchy horizon. Accordingly, from the equation  $f(r_+)=0$, we get
\be \label{horizonte}
1-\frac{2Mr_+^2}{(r_+^2+q^2)^{\frac{3}{2}}}=0\,,
\ee
which by using the Bekenstein-Hawking area-entropy relationship, $S = \pi r_+^2$,
can be rewritten as 
\be 
\label{inhomoge} 
M(s,q)=\frac{1}{2s}\left(
s+q^2\right)^{\frac{3}{2}}\,,
\ee
with $s=S/\pi$.
 This last equation relates all the
thermodynamic variables entering the black hole  metric in the
form of a fundamental thermodynamic equation $M = M(s, q)$.

\begin{figure}[h]
\includegraphics[scale=0.4]{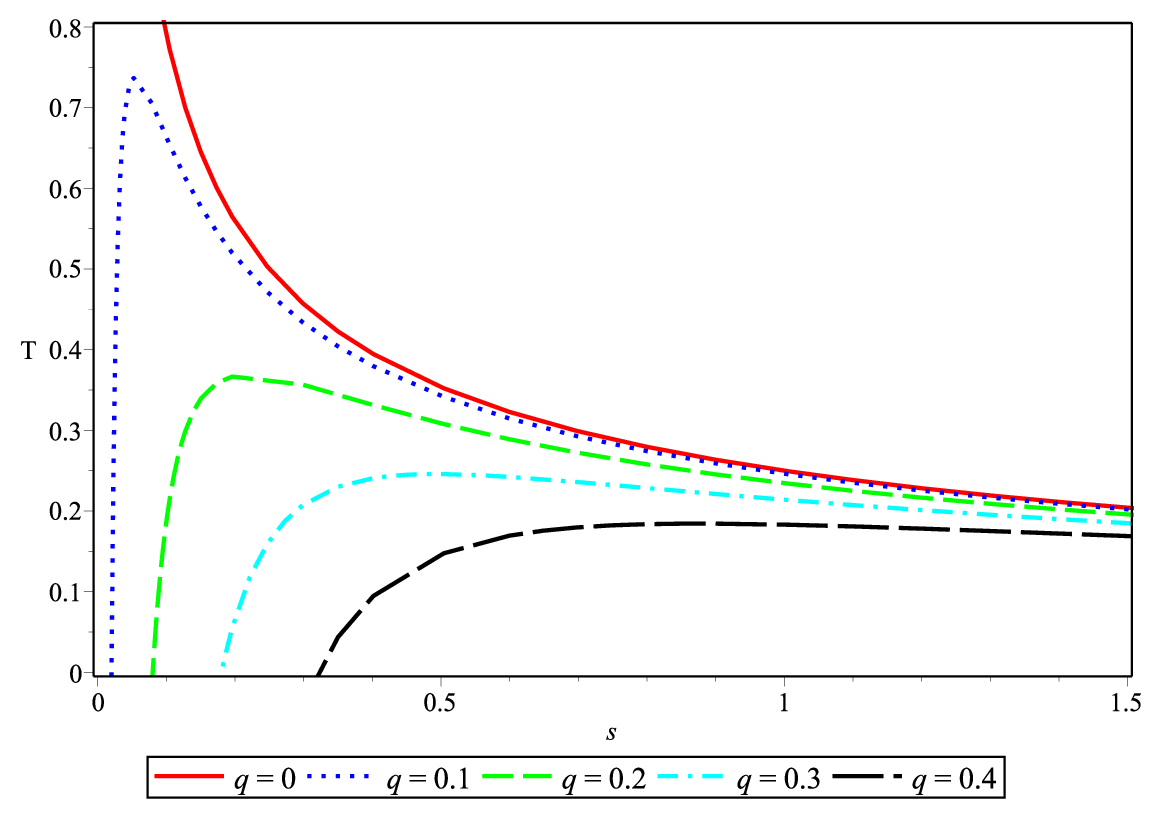} 
\caption{The
temperature $T$ as a function of the entropy $s$ for different values of the magnetic charge $q$. }
\label{fig1}
\end{figure}

\begin{figure}[h]
\includegraphics[scale=0.4]{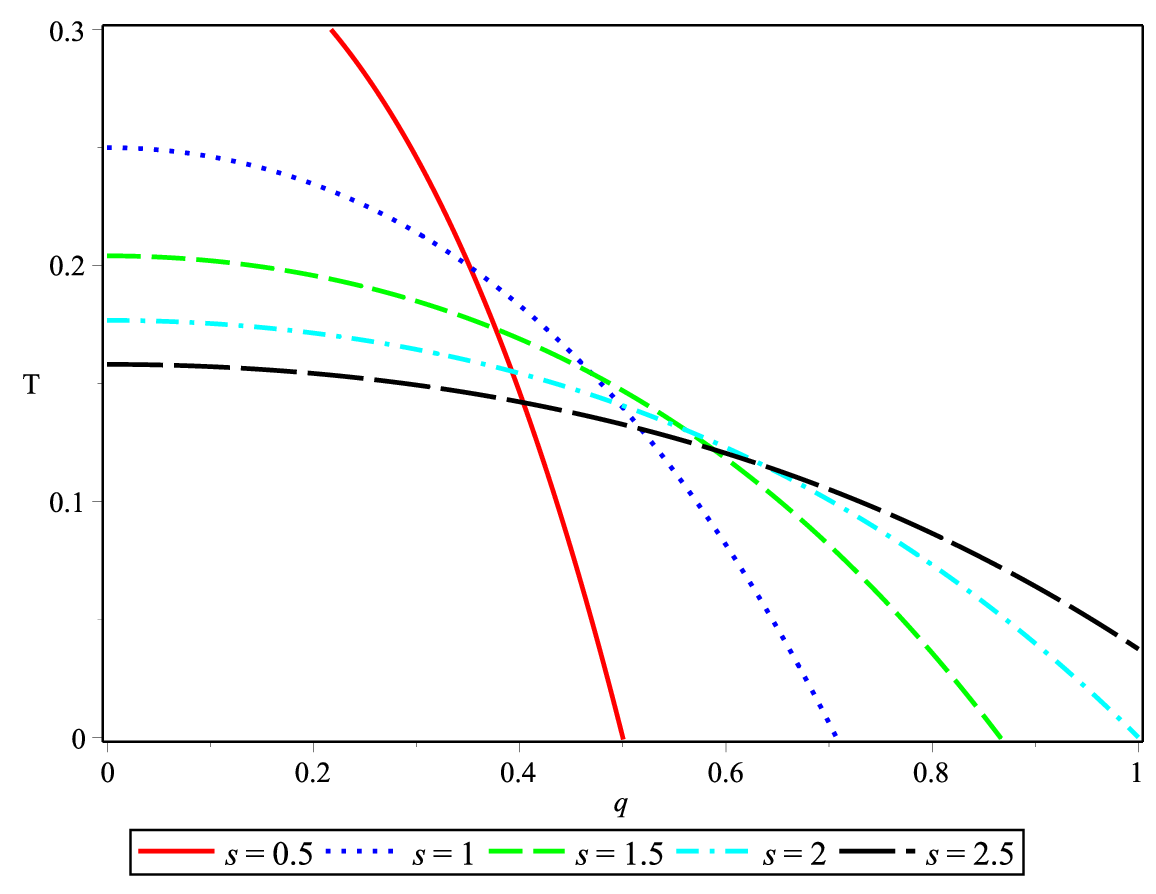} 
\caption{The
temperature $T$ as a function of the magnetic charge $q$ for different values of the entropy $s$.  }
\label{fig1b}
\end{figure}

Furthermore, the first law of  thermodynamics and
the thermodynamic equilibrium condition are given by the
expressions \cite{callen}
\be \label{firstlaw} dM=Tds+\phi dq\,, \ \ 
T =\Big(
\frac{\partial M}{\partial s}\Big)_{q} , \ \
\phi = \Big(
\frac{\partial M}{\partial q}\Big)_{s} ,
\ee
where $\phi$ is the variable dual to the magnetic charge $q$, which can be interpreted as the magnetic potential. For the fundamental equation (\ref{inhomoge}), we obtain  
\bea \label{ThermCond2} T &=& \frac{1}{4}\frac{(s-2q^2)\sqrt{q^2+s}}{s^2},\\
\label{ThermCond3}\phi  &=&
\frac{3}{2}\frac{q\sqrt{q^2+s}}{s}\,.
\eea
It is easy to show that the temperature (\ref{ThermCond2})
coincides with the Hawking temperature \cite{hawking1974black}. 
The expression for the temperature $T$ imposes a condition on the possible values of the entropy, namely, $s>2q^2$. A numerical analysis of the temperature shows that in the allowed interval it 
increases rapidly as a function of entropy $s$ until it reaches
its maximum value at $s=(2+\sqrt{12})q^2$. Then, as the entropy
increases, the temperature becomes a monotonically decreasing
function. This behavior is illustrated in Fig. \ref{fig1}. We see that the presence of the magnetic charge drastically changes the behavior of the temperature. Indeed, in the limiting Schwarzschild case $(q=0)$, the temperature diverges as the zero-entropy limit is approached. The magnetic charge eliminates this divergence and, instead, leads to a more physical behavior in which a zero-temperature limit is always reached for finite values of the entropy.  
In Fig.\ref{fig1b}, we plot the temperature in terms of the magnetic charge for fixed values of the entropy. The divergence that appears in the limit $s\rightarrow 0$ is now ``covered" by a region of negative temperature.  

\begin{figure}
\includegraphics[scale=0.4]{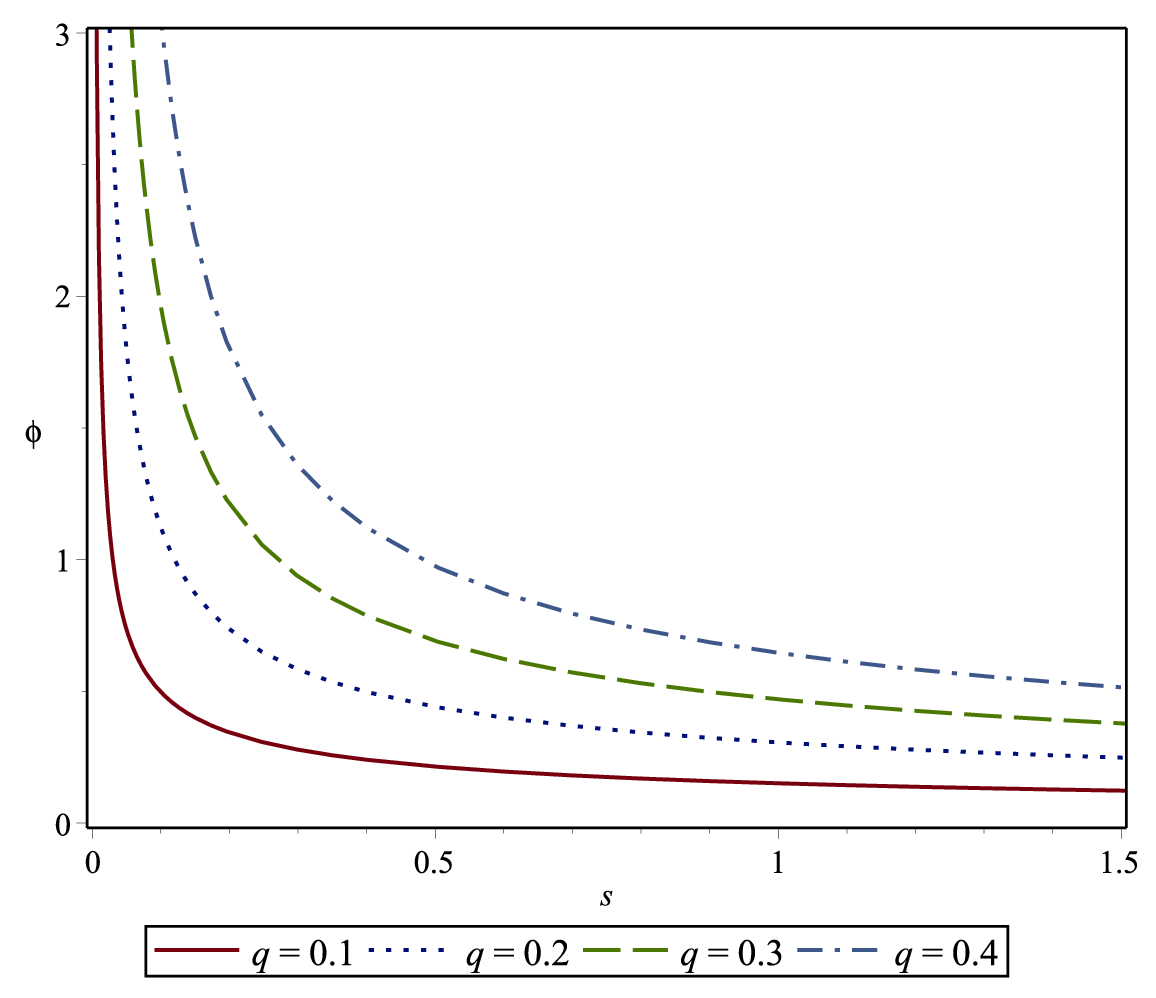}
    \caption{Behavior of the magnetic potential $\phi$ as a function of the entropy for different values of the magnetic charge $q$. }
    \label{fig2}
\end{figure}

On the other hand, the magnetic potential $\phi$ is well-defined for any positive values of the entropy and diverges as $s$ approaches the zero value. As the entropy increases, the magnetic potential becomes a monotonically decreasing function, which vanishes asymptotically (see Fig. \ref{fig2}).

\section{Geometrothermodynamics of the Bardeen  black hole} 
\label{sec:gtd}

As indicated above, the main goal  of GTD is to represent thermodynamic properties in terms of geometric concepts, i.e., GTD relates  geometric properties of the equilibrium space 
$\mathcal{E}$ with physical concepts like thermodynamic
interaction or phase transitions. In this context, curvature singularities
represent points of the manifold $\mathcal{E}$, where 
differential geometry cannot be applied anymore and, therefore, we
expect a similar behavior from a thermodynamic point of view. Indeed, equilibrium thermodynamics breaks down when the
system undergoes phase transitions.  This is an indication that  curvature
singularities can be interpreted as the geometric representation of phase
transitions. In general, singularities correspond to points of the
equilibrium space $\mathcal{E}$, where the laws of 
thermodynamics are no more valid, and 
denote locations where critical thermodynamic processes occur. In this section, we will apply these ideas in the framework of GTD  to investigate the properties of a regular black hole.

The first step consists in fixing the thermodynamic potential to be used in GTD. Usually, in classical thermodynamics, the entropy representation is preferred because it is appropriate to interpret the results of calculations. In the case of the Bardeen black hole, from Eq.(\ref{inhomoge}), we obtain 
\be
s= \left(\frac{1}{3}A^{1/3}+\frac{4}{3}\frac{M^2}{A^{1/3}}+\frac{2}{3}M\right)^2 - q^2,
\label{feqrup}
\ee
with 
\be
A= 8M^3 + 3q\sqrt{ 81M^2q^2-48M^4} - 27 Mq^2,
\ee
for the entropy representation. The use of this cumbersome expression leads to complicated results that make the analysis of the physical consequences difficult. Since the physical results in GTD are independent of the thermodynamic potential, we prefer to use the mass representation as defined by the fundamental equation $M=M(s,q)$, as given in Eq.(\ref{inhomoge}). From there it follows that the system possesses two thermodynamic degrees of freedom ($n=2$), the thermodynamic potential is given by the mass $M$ and the coordinates of the equilibrium space are $E^a=(s,q)$. Then, from Eqs. (\ref{gdownI})-(\ref{gdownIII}), we obtain the line elements
\bea \label{gdown2dI} g^{I}&=&\Big(\beta_s s \frac{\partial
M}{\partial s}+\beta_q q \frac{\partial M}{\partial
q}\Big)\Big(\frac{\partial^2 M}{\partial s^2 }ds^2
+2\frac{\partial^2 M}{\partial s \partial q }ds dq
+\frac{\partial^2 M}{\partial q^2 }dq^2\Big) \,,\\
g^{II}&=&\Big(\beta_s s \frac{\partial M}{\partial s}+\beta_q q
\frac{\partial M}{\partial q}\Big)\Big(-\frac{\partial^2
M}{\partial s^2 }ds^2
+\frac{\partial^2 M}{\partial q^2 }dq^2\Big) \,, \\
g^{III}&=&\beta_s s \frac{\partial M}{\partial s}\frac{\partial^2
M}{\partial s^2 } ds^2 +
\left(\beta_s s  
\frac{\partial M}{\partial s}
+\beta_q q \frac{\partial M}{\partial q}\right) 
\frac{\partial^2 M}{\partial s
\partial q} ds dq +\beta_q q \frac{\partial M}{\partial q
}\frac{\partial^2 M}{\partial q^2 }dq^2\label{gdown2dIII}\,,\eea 
where the specific values of the quasi-homogeneity coefficients $\beta_s$ and $\beta_q$ depend on the explicit form of the fundamental equation $M=M(s,q)$. Indeed, the fundamental equation (\ref{inhomoge})  is 
 a 
quasi-homogeneous function
i.e., 
\be M(\lambda^{\beta_s} s, \lambda^{\beta_q}
q)=\lambda^{\beta_M}M(s,q),
\ee
if the conditions 
\be
\beta_s=2\beta_M \quad {\rm and} \quad 
3\beta_q-\beta_s=\beta_M,
\label{qcond1}
\ee
i.e., 
\be 
\beta_s = 2 \beta_q,
\label{qcond2}
\ee
are satisfied. 
Taking into account the quasi-homogeneity property of the fundamental equation, the Euler identity reads \cite{quevedo3,qq23}
\be 
\beta_s s \frac{\partial M}{\partial s}
+\beta_q q
\frac{\partial M}{\partial q}=(3\beta_q-\beta_s)M ,
\label{euler}
\ee  
which can be shown to be identically satisfied by the function (\ref{inhomoge}). Consequently, the GTD metrics (\ref{gdown2dI})-
(\ref{gdown2dIII}) can be reduced to 
\bea \label{gdownredI} 
g^{I}&=&\beta_M M 
\Big(\frac{\partial^2 M}{\partial s^2 }ds^2
+2\frac{\partial^2 M}{\partial s \partial q }ds dq
+\frac{\partial^2 M}{\partial q^2 }dq^2\Big) \,,\\
g^{II}&=&\beta_M M \Big(-\frac{\partial^2
M}{\partial s^2 }ds^2
+\frac{\partial^2 M}{\partial q^2 }dq^2\Big) \,, \\
g^{III}&=&\beta_s s \frac{\partial M}{\partial s}\frac{\partial^2
M}{\partial s^2 } ds^2 +
\beta_M M  
\frac{\partial^2 M}{\partial s
\partial q} ds dq +\beta_q q \frac{\partial M}{\partial q
}\frac{\partial^2 M}{\partial q^2 }dq^2\label{gdownredIII}\,.
\eea 
Finally, considering the fundamental equation (\ref{inhomoge}), we obtain the explicit form of the metric components, which can be represented as follows
\be
g^I = \frac{\beta_M M}{2 s \sqrt{s+q^2}}\left[
\frac{8q^4+4sq^2-s^2}{4s^2}ds^2 -\frac{3q(s+2q^2)}{s}ds dq
+3(s+2q^2) dq^2 \right],
\ee
\be
g^{II} = \frac{\beta_M M}{2s\sqrt{s+q^2}}\left[
-\frac{8q^4+4sq^2-s^2}{4s^2}ds^2 +3(s+2q^2) dq^2
\right ],
\ee
\be
g^{III} = \frac{\beta_M}{4s^2}\left[ 
\frac{(s-2q^2)(8q^4+4sq^2-s^2)}{8s^2}ds^2
+3q\sqrt{s+2q^2}\left( M ds + 3q\sqrt{s+2q^2}dq \right)dq  
\right].
\ee
Notice that, after applying the relationships (\ref{qcond1}) and (\ref{qcond2}),  the explicit form of the metrics depends on the quasi-homogeneity coefficient $\beta_M$ only. This dependence, however, is represented by a constant conformal factor that does not affect the geometric properties of the metrics. This shows the generality of the GTD approach in the sense that it can be applied to homogeneous and quasi-homogeneous systems in the same manner.

\begin{figure}
    \centering
\includegraphics[scale=0.5]{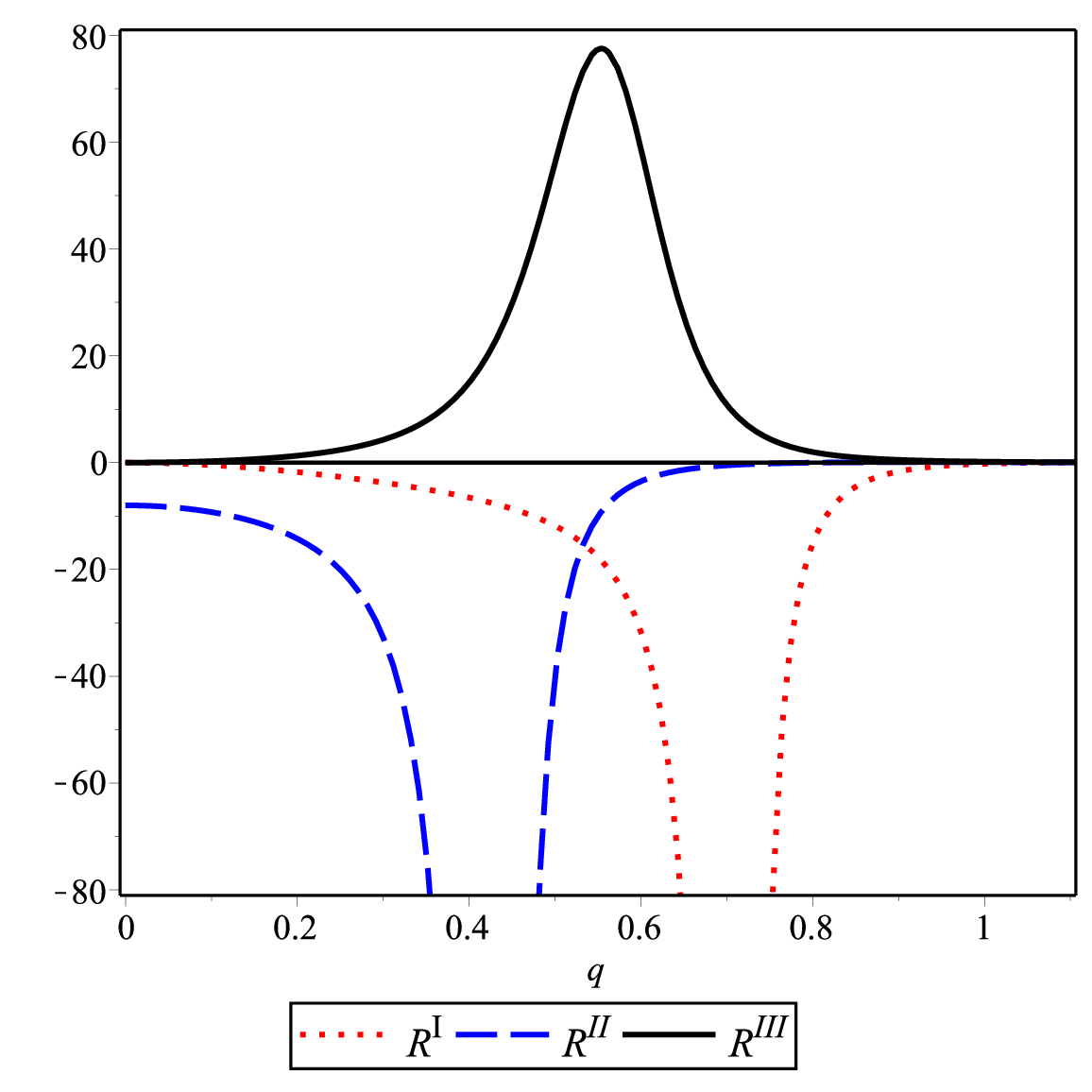}
    \caption{Behavior of the curvature scalars of the equilibrium space of the Bardeen black hole in terms of the parameter $q$. Here, we set $\beta_M=1$ and $s=1$.}
    \label{fig5a}
\end{figure}

A straightforward computation shows that the curvature scalars of the above metrics can be expressed as
\be
{R}^{I}={\frac {16\, {s}^{3}{q}^{2} \left( 4\,{q}^{4}-2\,s{q}^{2}-3\,{s}^{2}
 \right) }{\beta_M 
 (s^2-4q^ 4)^2 (s+q^2)^3 }}
 , \label{RI}
\ee
\be
R^{II}=
  {\frac { 8\, {s}^{3} \left( 32\,{q}^{6}-12\,{s}^{2}{q}^{2}-{s}^{3}
 \right) }{ \beta_M\left( s+{q}^{2} \right)  \left( -{s}^{2}+4\,s{q}^{2}+8\,{
q}^{4} \right) ^{2} \left( s+2\,{q}^{2} \right) ^{2}}} ,
\label{RII}
\ee
\be
R^{III} = {\frac {128\,{q}^{2}{s}^{3} \left( 12\,{q}^{4}-4\,s{q}^{2}+5\,{s}^{2}
 \right) }{\beta_M \left( s+2\,{q}^{2} \right)  \left( 5\,{s}^{3}-20\,{s}^{2}
{q}^{2}+66\,{q}^{6}+5\,s{q}^{4} \right) ^{2}}},
\label{RIII}
\ee
respectively. These results show that the equilibrium space of the Bardeen regular black hole is in general curved, indicating the existence of thermodynamic interaction,  and that each of the above metrics describes different aspects of the thermodynamic interaction. In Fig. \ref{fig5a}, we show the behavior of the curvature scalars corresponding to each metric in terms of the values of the magnetic charge $q$. We see that $R^I$ and $R^{II}$ become singular in some particular locations, whereas the scalar $R^{III}$ is regular everywhere. 
The interpretation of the curvature singularities in terms of the thermodynamic properties of the system represents one of the main goals of GTD.

Furthermore, we emphasize that the advantage of the above reduced representation of the GTD metrics is that by using the Euler identity (\ref{euler}) in the corresponding curvature scalars, it can be shown that the singularities are determined by the conditions \cite{qq23}
\be
I: \frac{\partial^2
M}{\partial s^2 } \frac{\partial^2
M}{\partial q^2 } -\left(\frac{\partial^2
M}{\partial s \partial q }\right)^2 =0 \ ,
\label{condI}
\ee
\be
II: \frac{\partial^2
M}{\partial s^2 }\frac{\partial^2
M}{\partial q^2 } = 0,
\ee
\be
III: \frac{\partial^2
M}{\partial s \partial q  }  = 0 ,
\ee
which exactly represent the locations where the stability conditions of a thermodynamic system with two degrees of freedom are broken \cite{callen}.
Note that these conditions are not completely independent. For instance, if condition $II$ is satisfied then conditions $I$ and $II$ are identical. In the same way, if condition $II$ is satisfied, conditions $I$ and $III$ coincide. 
 Furthermore, from the fundamental equation (\ref{inhomoge}), we obtain
\be
I: \frac{3}{16} \frac{s^2-4q^4}{s^4} = 0,
\ee
\be
II: \frac{3}{16}\,{\frac { \left( -{s}^{2}+4\,s{q}^{2}+8\,{q}^{4} \right)  \left( 
2\,{q}^{2}+s \right) }{ \left( s+{q}^{2} \right) {s}^{4}}}
=0,
\ee
\be
III: -\frac{3}{4}\,{\frac {q \left( 2\,{q}^{2}+s \right) }{\sqrt {s+{q}^{2}}{s}^{2}
}}
=0.
\label{condIII}
\ee
As expected, these conditions coincide with the locations of the curvature singularities of the GTD metrics given in Eqs.(\ref{RI})--(\ref{RIII}).
We see that there are only two types of curvature singularities determined by the conditions
\be
I: s=2q^2\ ,
\label{singI}
\ee
\be
II: -{s}^{2}+4\,s{q}^{2}+8\,{q}^{4} = 0 \ .
\label{singII}
\ee
This completes the analysis of the curvature of the GTD metrics for the Bardeen black hole. The conditions (\ref{singI}) and (\ref{singII}) determine the locations where the GTD approach breaks down, which, as we will show below, correspond to values of the independent thermodynamic variables at which the approach of equilibrium thermodynamics cannot be applied anymore.

The physical interpretation of curvature singularities in GTD has been extensively discussed in the literature. In particular, Tharanath et al. \cite{Tharanath2015} showed explicitly that, for several regular black-hole configurations that include the Bardeen spacetime, the divergences of the heat capacity occur exactly at the points where the GTD scalar curvature becomes singular. This correspondence strongly supports the interpretation of GTD curvature singularities as indicators of thermodynamic phase transitions. 
In this work, instead, we perform a detailed analysis of the thermodynamic and geometrothermodynamic properties of the Bardeen black hole, including the investigation of the critical behavior encoded in the GTD. 
Consequently, the singular structure obtained in the present analysis can be interpreted as encoding genuine thermodynamic instabilities of the black-hole system.



\section{Thermodynamic curvature and phase transitions}
\label{sec:inter}

To interpret the geometric results obtained from the analysis of the equilibrium space and its metrics,  we should compare them with the results derived from classical black hole  thermodynamics. In particular, the response functions of a thermodynamic system
are used to indicate the presence of second-order phase
transitions. In the case of a thermodynamic system with two
degrees of freedom, there exist only three independent response
functions \cite{callen}, one of them being the heat capacity. 
On the other hand, according to the standard construction of black hole thermodynamics \cite{davies}, the phase transitions of a black hole are determined by the behavior of the heat capacity. 
For the Bardeen spacetime, the heat capacity at constant $q$ is given by
\cite{callen}
\bea \label{capacity} C_{q}=T\Bigg(\frac{\partial s}{\partial
T}\Bigg)_{q}=\Bigg(\frac{\frac{\partial M}{\partial
s}}{\frac{\partial^2 M}{\partial s^2}}\Bigg)_{q}\,,\eea 
where the subscript indicates that the derivatives are calculated keeping the
magnetic charge constant. Using the fundamental equation
(\ref{inhomoge}),  we get
\bea
\label{heatcapacity}
C_{q}=\frac{2s(s-2q^2)(q^2+s)}{8q^4+4q^2s-s^2}\,.
\eea

\begin{figure}[h]
\includegraphics[scale=0.35]{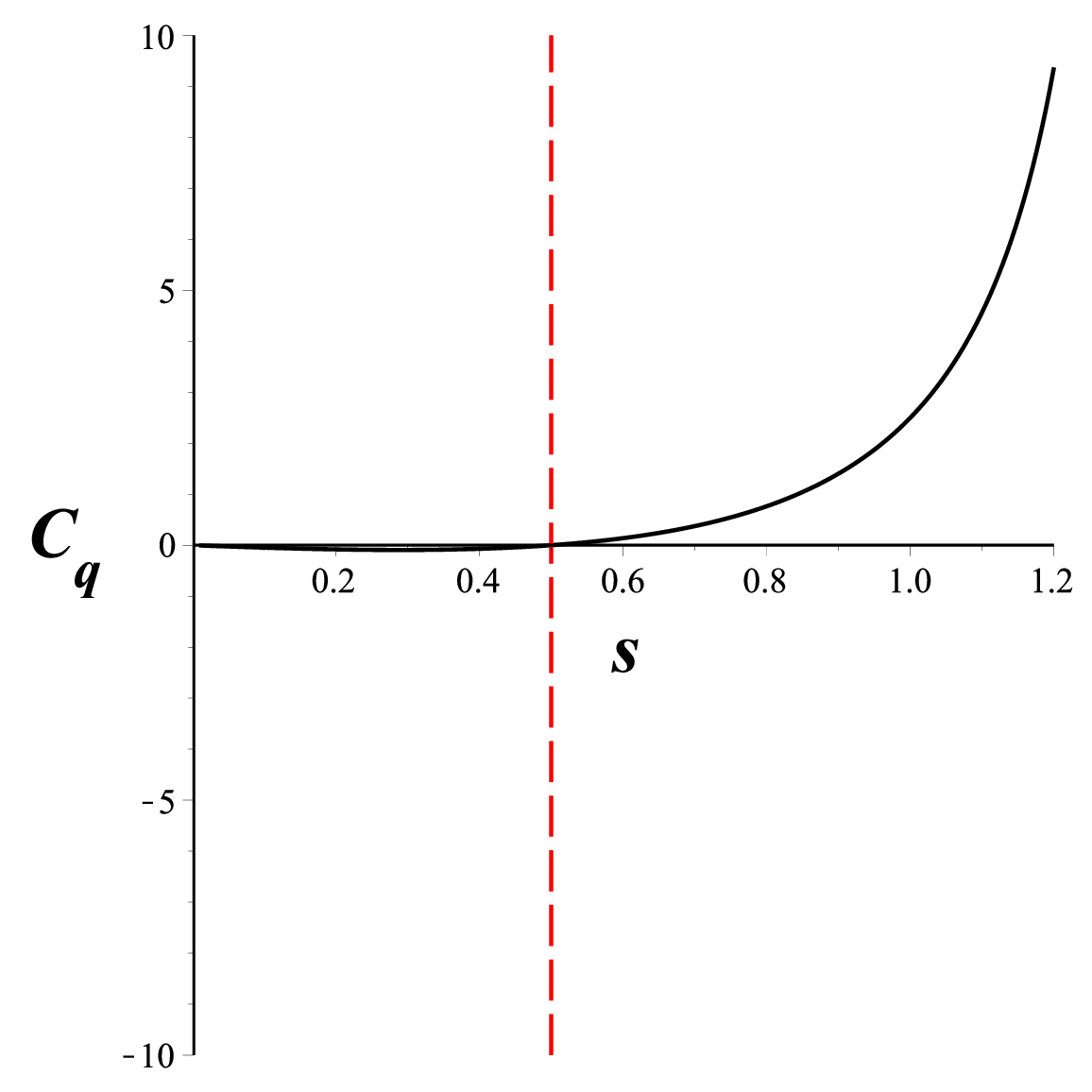}
\includegraphics[scale=0.4]{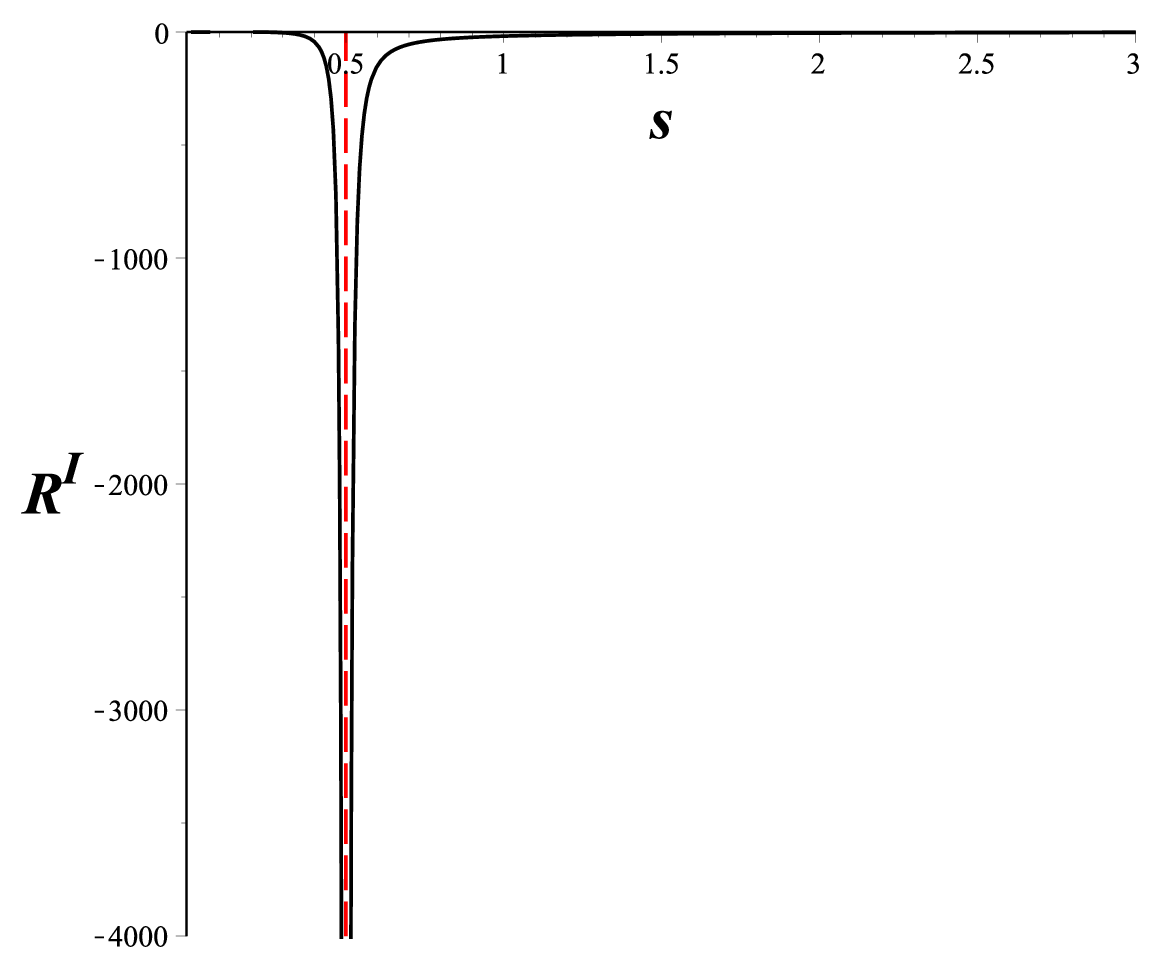}
\caption{Left panel: The heat capacity $C_{q}$ as a function of the entropy
$s$. Right panel: The curvature scalar of the metric  $g^I$ as a function of the entropy
$s$. We have considered in both cases $q=\frac{1}{2}$ and
$\beta_q=1$. The zero of the heat capacity coincides with the curvature singularity of $R^I.$ }
\label{fig4}
\end{figure}

The vanishing of the heat capacity is usually accompanied by a
phase transition during which the stability properties of the system are modified.  In the case of the regular Bardeen black hole the heat
capacity (\ref{heatcapacity}) is zero at the point $s=2q^2$. 
On the other hand, these points correspond to the singularity of
the curvature scalar (\ref{singI}). Therefore, this result shows
that the GTD line element $g^I$  describes from a geometric point of view the stability thermodynamic properties 
of the regular Bardeen black hole. In fact, the general relationship (\ref{condI}) is known in classical thermodynamics to represent the stability condition for a system with two degrees of freedom \cite{callen}. When the stability condition is not satisfied, the system undergoes a phase transition.
In Fig. \ref{fig4}, we show that, in fact, for the Bardeen black hole the zero of the heat capacity corresponds to a singularity in the curvature of the metric $g^I.$

\begin{figure}[h]
\includegraphics[scale=0.35]{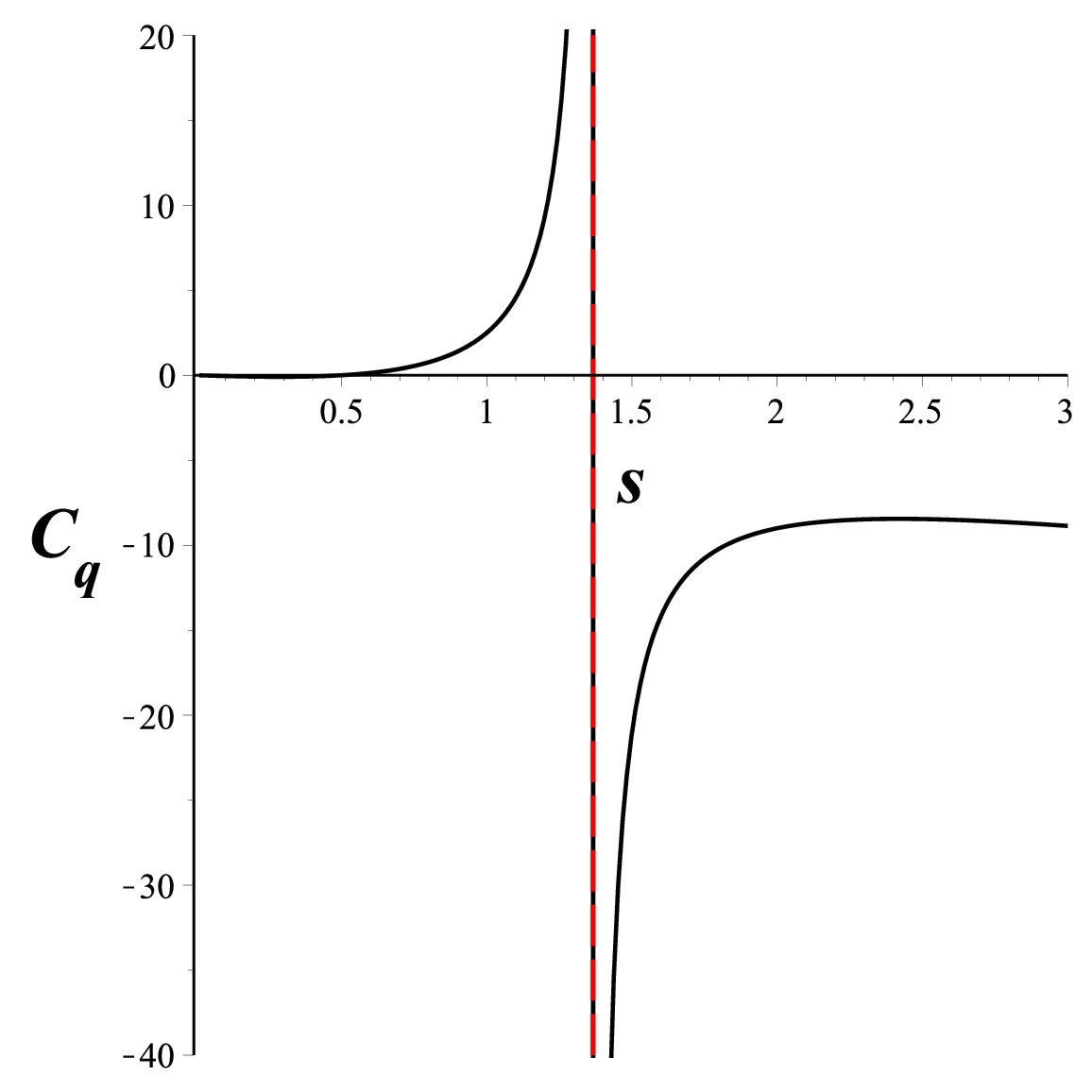}\includegraphics[scale=0.4]{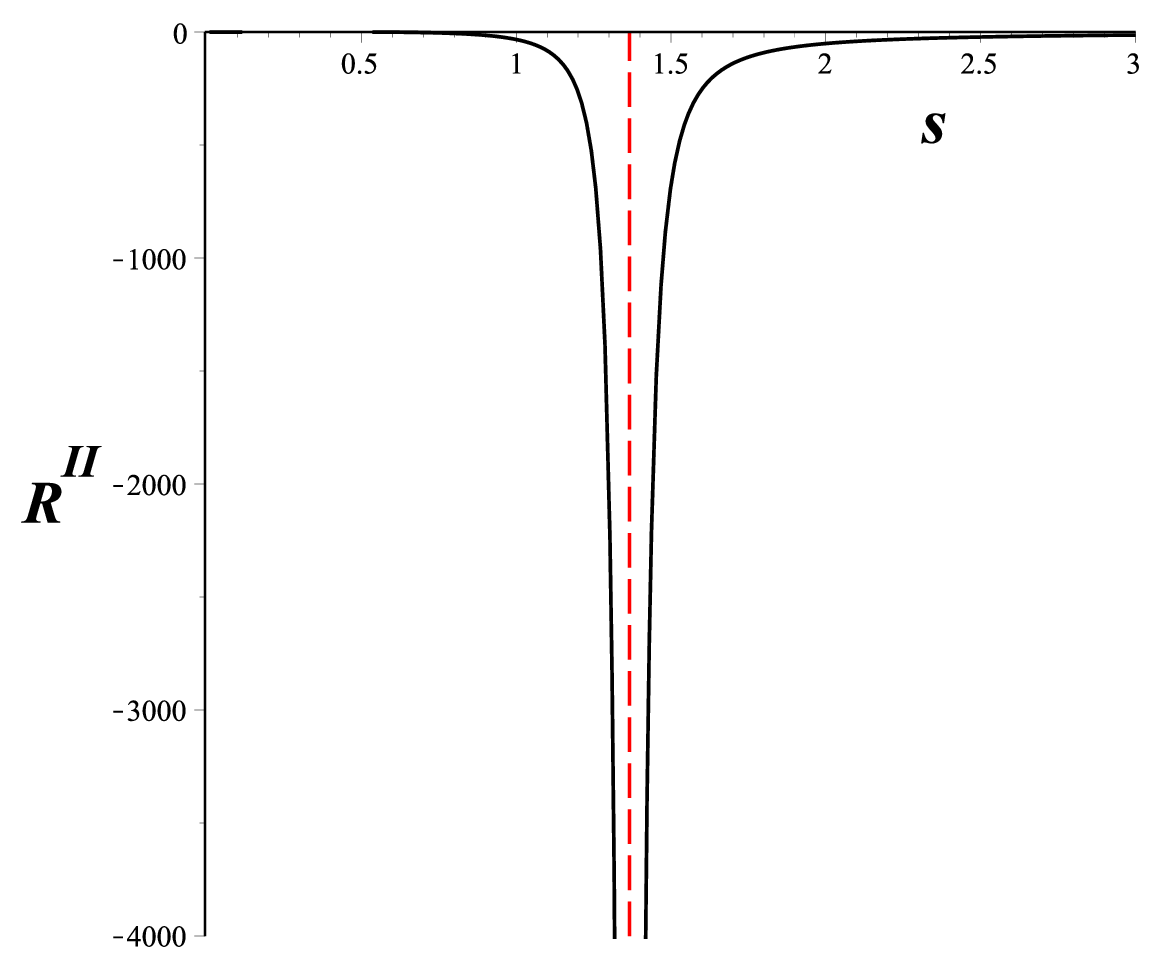}
\caption{Left panel: The heat capacity $C_{q}$ as a function of the entropy
$s$. Right panel: The curvature scalar of the metric $g^{II}$ as a function of the entropy
$s$. Here, $q=\frac{1}{2}$ and
$\beta_q=1$. The second-order phase transition occurs at the divergence of the heat capacity, which coincides with the curvature singularity of the metric $g^{II}_{ab}.$ 
}
\label{fig5}
\end{figure}

Furthermore, according to Davies \cite{davies}, second-order phase transitions
take place at those points where the heat capacity diverges, i.e., according to Eq.(\ref{heatcapacity}), at 
\be 
\label{sing} 8q^4+4q^2s-s^2=0\,,
\ee
an equation with the non-trivial solutions $q=\pm \frac{1}{2}\sqrt{(\sqrt{3}-1)s}$. 
We observe immediately that this condition is
identical to the condition for the existence of curvature
singularities in the equilibrium space of the metric $g^{II}_{ab}$, as
given in the expression (\ref{singII}). In Fig. \ref{fig5},  
 we compare 
the behavior of the heat capacity $C_q$ and the scalar curvature $R^{II}$.

These results show that there exist curvature
singularities in the equilibrium space described by the line elements $g^I$ and $g^{II}$ at those points where phase transitions occur in the Bardeen black hole. 
Furthermore, as follows from condition (\ref{condIII}), the curvature scalar of the metric $g^{III}$ is
regular for the allowed physical range of the variables $s$ and $q$. 

\begin{figure}
    \centering
    \includegraphics[scale=0.35]{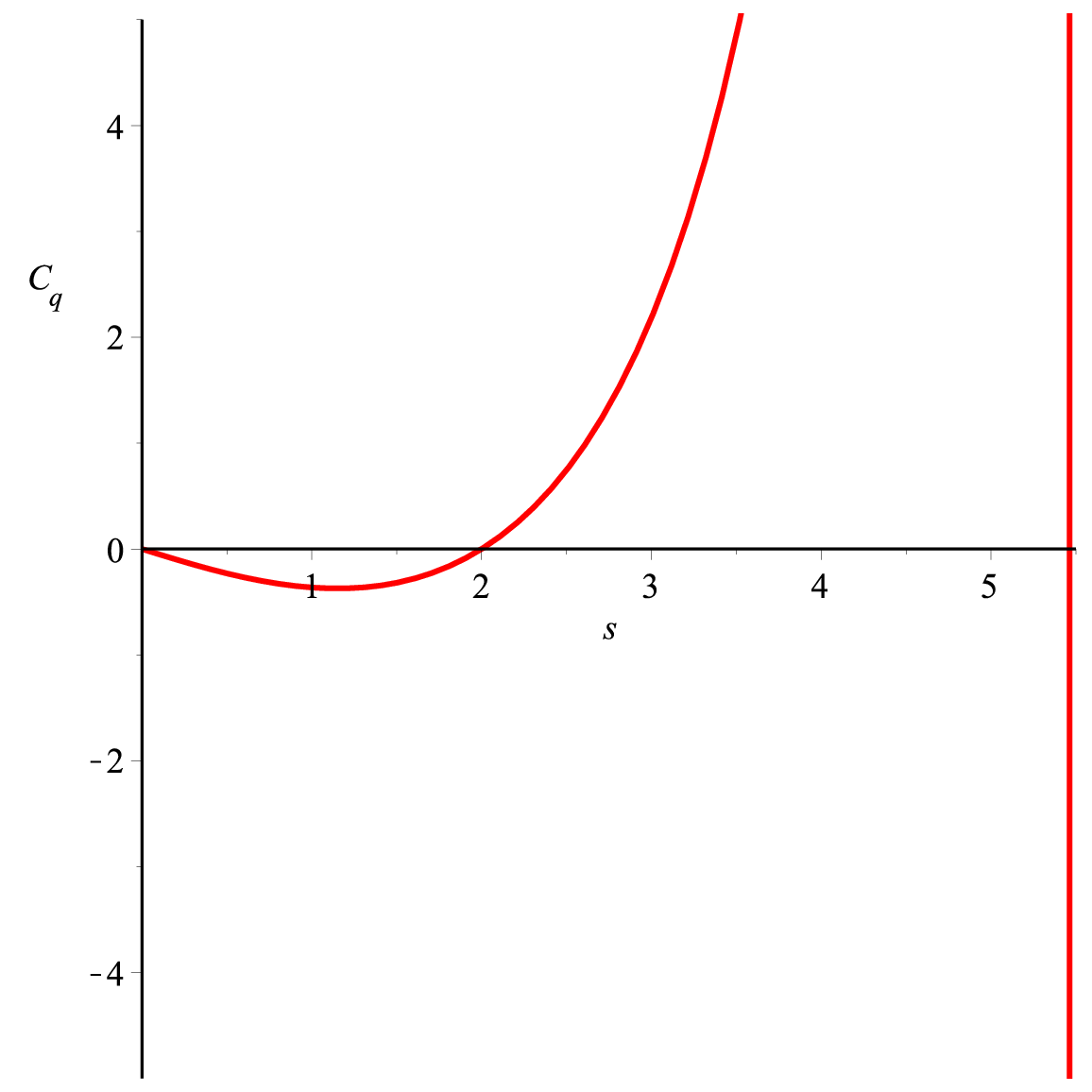}
       \includegraphics[scale=0.35]{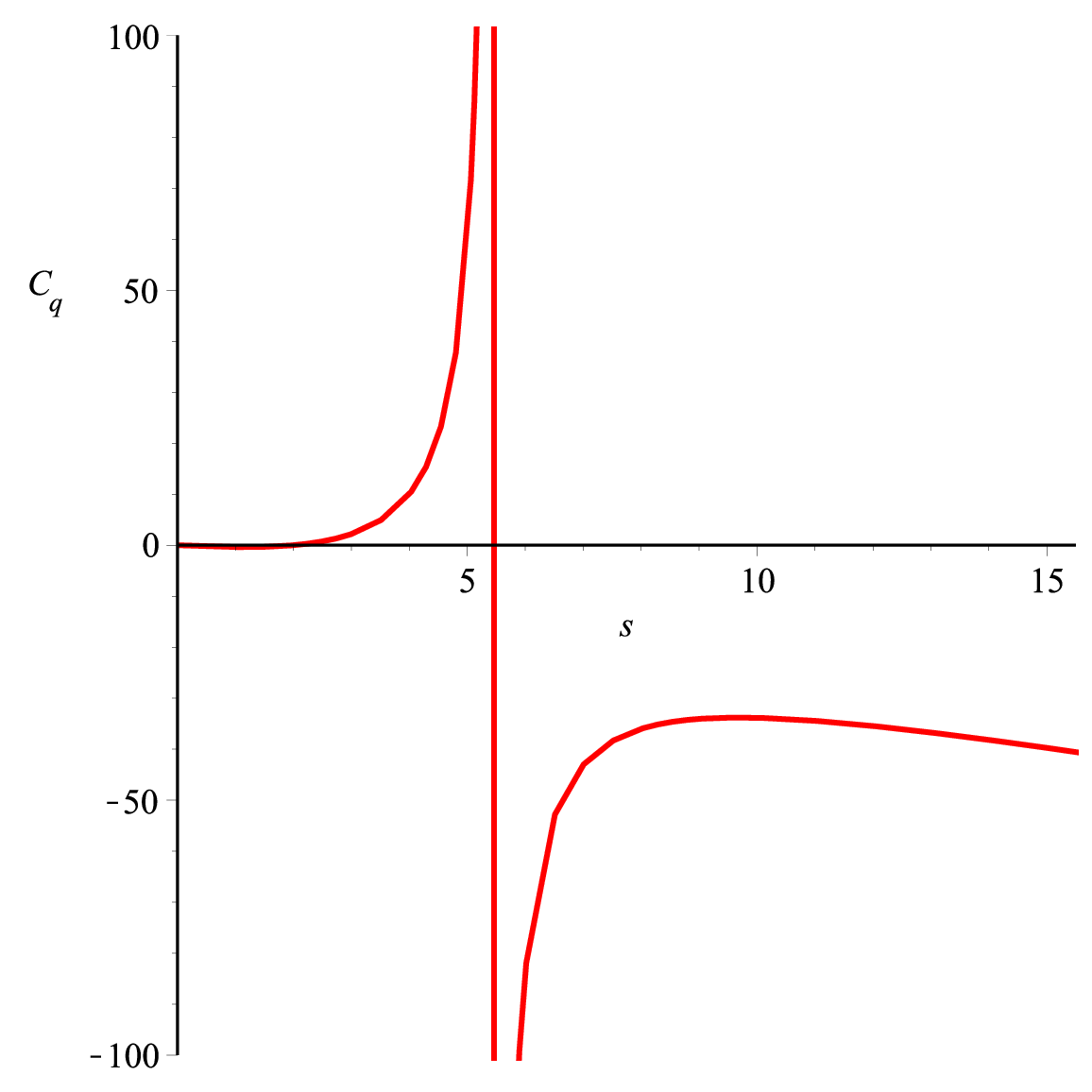}
       \caption{Behavior of the heat capacity for a fixed magnetic charge $(q=1)$ in terms of the entropy. The left panel emphasizes the interval with negative values in the heat capacity before the phase transition. }
    \label{fig3}
\end{figure}

From the above analysis, we can derive the 
 complete phase transition structure of the Bardeen black hole as follows. 
In Fig. \ref{fig3}, we illustrate the behavior of the heat capacity as a function of the entropy. First, the black hole is unstable for small values of the entropy. Then, as the entropy increases, it becomes stable in the interval (see Fig. \ref{fig3}, left panel)
\be  
2q^2<s<\frac{4}{\sqrt{3}-1}.
\ee
 The divergence at $s=\frac{4}{\sqrt{3}-1}$ corresponds to a second-order phase transition, which brings the black hole back to an unstable state (see, Fig. \ref{fig3}, right panel).

\begin{figure}
    \centering
    \includegraphics[scale=0.4]{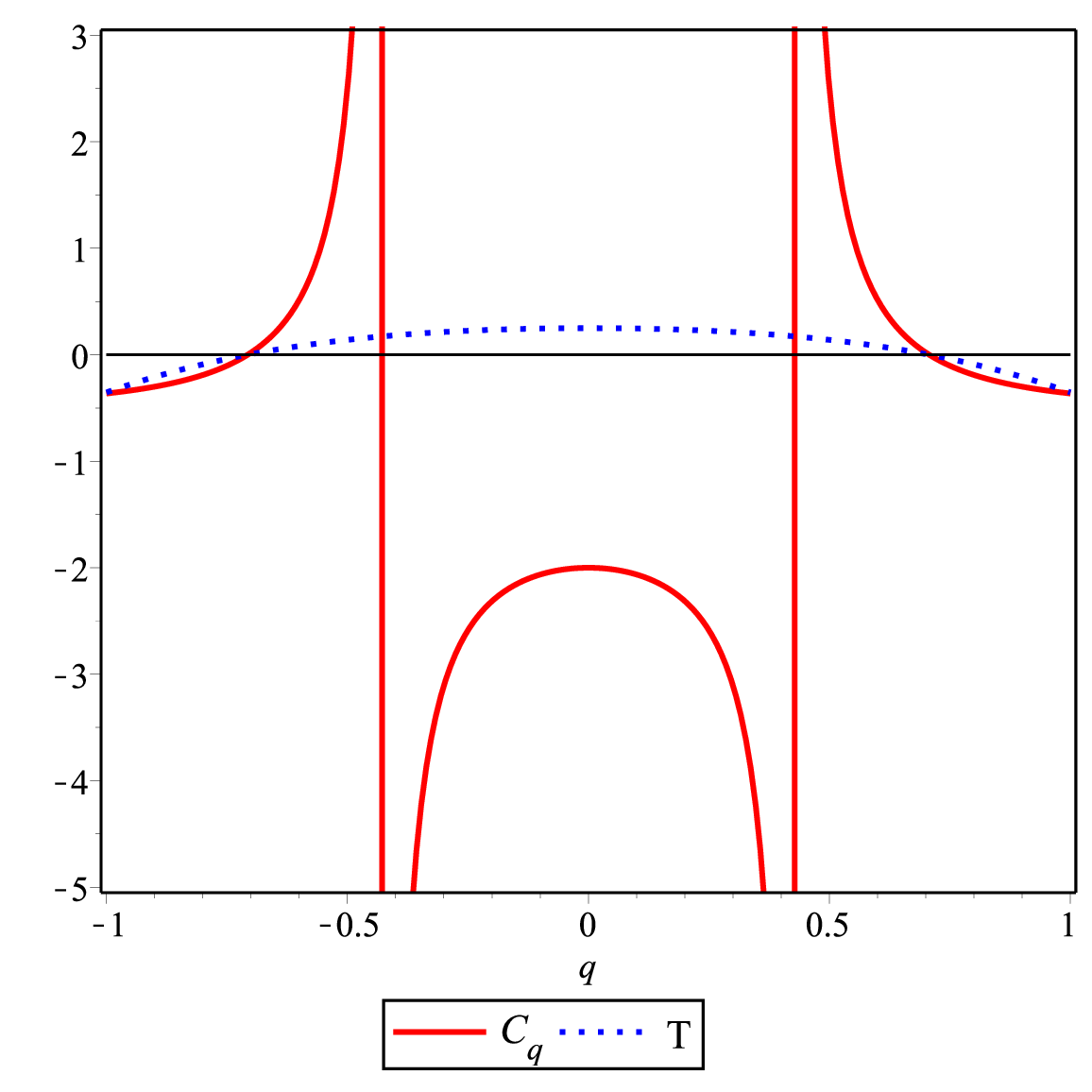}
    \caption{Behavior of the heat capacity as a function of the magnetic charge. Here, we fix $s=1$ for concreteness. The black hole can exist only in the region of positive temperature values and is stable only in the small interval with positive values of the heat capacity.
    }
    \label{fig4new}
\end{figure}

In Fig. \ref{fig4new}, we plot the heat capacity in terms of the magnetic charge. The black hole is unstable in the interval 
\be -\frac{1}{2}\sqrt{(\sqrt{3}-1)s} < q < \frac{1}{2}\sqrt{(\sqrt{3}-1)s},
\ee
including the uncharged case, which is well-known to be unstable. The phase transition occurring at $q=\pm 
\frac{1}{2}\sqrt{(\sqrt{3}-1)s}$ brings the system into a stable state, which is the starting state of the stable intervals 

\be
-\sqrt{\frac{s}{2}}<q<-\frac{1}{2}\sqrt{(\sqrt{3}-1)s}, \ \ {\rm  and} \ \ 
\frac{1}{2}\sqrt{(\sqrt{3}-1)s}<q<\sqrt{\frac{s}{2}}.
\ee
A new phase transition occurs at $q=\pm \sqrt{\frac{s}{2}} $, which brings the black hole back to an unstable state. However, this transition is not allowed because it corresponds to a zero-temperature, which is positive definite in the interval $-\sqrt{\frac{s}{2}} < q +\sqrt{\frac{s}{2}}$ only,  as illustrated in Fig. \ref{fig4new}.

To conclude this section, we will compare our results with those obtained in thermodynamic geometry with the use of Hessian metrics of particular thermodynamic potentials. The use of the entropy as thermodynamic potential turns out to be difficult to be computed because of the functional complexity of the corresponding expression (\ref{feqrup}). Instead, we used the mass function (\ref{inhomoge}) as the thermodynamic potential for the Weinhold metric \cite{Weinhold} 
\be
g^W_{ab} = \frac{\partial ^2 M}{\partial E^a \partial E^b}\ , \quad E^a=(s,q)
\ee
can be carried out easily. Indeed, a direct computation shows that 
\be
g^W = \frac{1}{2 s \sqrt{s+q^2}}\left[
\frac{8q^4+4sq^2-s^2}{4s^2}ds^2 -\frac{3q(s+2q^2)}{s}ds dq
+3(s+2q^2) dq^2 \right],
\ee
from which we obtain the curvature scalar
\be
R^W =
-{\frac {2\,{s}^{3} \left( s+4\,{q}^{2} \right) }{ \left( s+{q}^{2}
 \right) ^{3/2} \left( s+2\,{q}^{2} \right)  \left( -s+2\,{q}^{2}
 \right) ^{2}}}. 
\ee
It follows that the only curvature singularity occurs at $s=2q^2$, which coincides with the singularity of the metric $g^I$  given in Eq.(\ref{singI}) and associated with the zero-temperature limit of the black hole. However, the GTD singularity (\ref{singII}) is not predicted by the Weinhold metric. We conclude that in this case the thermodynamic geometry associated with the Weinhold metric does not reproduce the Davies phase transition of the Bardeen black hole. 

\section{Critical exponents and scaling behavior}
\label{sec:cri}

The previous section shows that the heat capacity and the curvature scalar
associated with the metric $g^{II}$ become singular at the critical point, where  the Davies phase
transition takes place. We now investigate the behavior of these quantities in the
neighborhood of the critical point.

The heat capacity at constant magnetic charge is given in Eq.(\ref{heatcapacity}).
Its divergence occurs when
$ 8q^4+4q^2s-s^2=0$. 
In terms of the entropy, the only positive solution is given by 
\be
s_c = 2(1+\sqrt{3})q^2. 
\ee
To analyze the behavior near the critical point, we introduce the quantity
\begin{equation}
\Delta s=s-s_c.
\end{equation}
Then, the denominator of the heat capacity can be expanded as
$ 
8q^4+4q^2s-s^2
=
-4\sqrt{3}q^2\Delta s
-(\Delta s)^2$.
Since the numerator of $C_q$ does not vanish at $s=s_c$, the leading
behavior of the heat capacity is
\begin{equation}
C_q=
-\frac{2(9+5\sqrt{3})q^4}{\Delta s}
+\mathcal{O}(1),
\label{coeffCq}
\end{equation}
implying that 
$
|C_q|\sim |\Delta s|^{-1}.
$ 

On the other hand, the temperature of the Bardeen black hole was derived in Sec. \ref{sec:bardeen} and can be expressed as 
\begin{equation}
T=
\frac{(s-2q^2)\sqrt{s+q^2}}{4s^2}.
\end{equation}
At the critical point, the first derivative of the temperature vanishes, $ 
\left(\frac{\partial T}{\partial s}\right)_{s=s_c}=0,
$ 
showing that the critical entropy corresponds to the maximum of the
temperature. Moreover, the critical temperature is
\begin{equation}
T_c=T(s_c,q)
=
\frac{\sqrt{9+6\sqrt{3}}}
{16(2+\sqrt{3})q},
\end{equation}
and the expansion of the temperature around the critical point therefore starts
at second order:
\begin{equation}
T-T_c=
\frac{1}{2}
\left(\frac{\partial^2T}{\partial s^2}\right)_{s=s_c}
(\Delta s)^2
+
\mathcal{O}\left((\Delta s)^3\right).
\end{equation}
Since the second derivative is negative, this relation can be written as
\begin{equation}
T_c-T=
T_c\frac{9-5\sqrt{3}}{24q^4}
(\Delta s)^2
+
\mathcal{O}\left((\Delta s)^3\right).
\end{equation}

Introducing the reduced temperature
$ 
t=\frac{T_c-T}{T_c},
$ 
we obtain
\begin{equation}
t=
\frac{9-5\sqrt{3}}{24q^4}
(\Delta s)^2
+
\mathcal{O}\left((\Delta s)^3\right).
\end{equation}
Consequently,
$ 
|\Delta s|\sim t^{1/2}.
$ 
Using the leading behavior of the heat capacity found above, it follows that
\begin{equation}
|C_q|\sim t^{-1/2}
\sim |T-T_c|^{-1/2}.
\end{equation}
Therefore, if the critical exponent $\alpha$ is defined by
$ 
|C_q|\sim |T-T_c|^{-\alpha},
$ 
we obtain $\alpha = \frac{1}{2}$. 


We now analyze the curvature scalar associated with the Legendre invariant
metric $g^{II}$. For the Bardeen fundamental equation, a direct calculation
gives
\begin{equation}
R^{II}
=
-\frac{
8s^3\left(32q^6-12q^2s^2-s^3\right)
}{
\beta_M(s+q^2)(s+2q^2)^2
\left(8q^4+4q^2s-s^2\right)^2
}.
\end{equation}
The singular factor in the denominator is the same one that determines the
divergence of the heat capacity,
$ 
D(s,q)=8q^4+4q^2s-s^2.
$ 
Near the critical point, it behaves as 
$
D(s,q)
=
-4\sqrt{3}q^2\Delta s
-(\Delta s)^2,
$ 
and therefore
$ D(s,q)^2
=
48q^4(\Delta s)^2
+
\mathcal{O}\left((\Delta s)^3\right).
$ 
The remaining factors in $R^{II}$ are finite and nonzero at $s=s_c$.
Consequently, the leading behavior of the curvature scalar is
\begin{equation}
R^{II}
=
\frac{64\sqrt{3}}{3\beta_M}
\frac{q^2}{(\Delta s)^2}
+
\mathcal{O}\left(\frac{1}{\Delta s}\right).
\label{coeffR}
\end{equation}
Thus,
$ 
R^{II}\sim |\Delta s|^{-2}
$, 
and using the relation $|\Delta s|\sim t^{1/2}$ obtained above, we find
\begin{equation}
|R^{II}|
\sim
t^{-1}
\sim
|T-T_c|^{-1}.
\end{equation}
If the curvature critical exponent $\gamma$ is defined by
$
|R^{II}|
\sim
|T-T_c|^{-\gamma},
$ 
then $\gamma=1$.

Furthermore, combining the expressions for the critical behavior of the heat capacity (\ref{coeffCq}) and the curvature $R^{II}$ (\ref{coeffR}), 
we obtain
the scaling relation
\begin{equation}
{
R^{II}
=
\frac{8(26\sqrt{3}-45)}
{9\beta_Mq^6}
C_q^2
+
\mathcal{O}(C_q)
}.
\end{equation}
Thus, close to the Davies transition,
$ {R^{II}\propto C_q^2}.$  This result shows that 
the curvature of the metric
$g^{II}$ contains information not only about the location of the phase
transition but also about the strength of the critical divergence. Applying a similar procedure, it can Consequently, thown that the curvature scalars $R^I$ and $R^{III}$ remain finite at the Davies transition. 
Consequently, the behavior of the three curvature scalars at the Davies point can,
therefore, be summarized as follows
\begin{equation}
R^{I}(s_c,q)<\infty,
\qquad
R^{II}\sim |T-T_c|^{-1},
\qquad
R^{III}(s_c,q)<\infty.
\end{equation}
Thus, among the three Legendre invariant metrics considered here, only
$g^{II}$ is enough to reproduce the critical behavior of the Davies phase transition.
The metric $g^{I}$ describes the extremal or stability boundary, whereas
$g^{III}$ remains regular throughout the physical domain.

In summary, the heat capacity of the Bardeen black hole diverges with the critical exponent $\alpha=1/2$, whereas the thermodynamic curvature associated with the GTD metric $g^{II}$ diverges with the exponent $\gamma=1$,  leading to the  scaling relation $R^{II}\propto C_q^{\,2}$. This result shows that the thermodynamic curvature contains information not only about the location of the Davies phase transition but also about its critical behavior. These results strengthen the geometric interpretation of GTD.

\section{Conclusions} \label{sec:con}
In this work, we have analyzed the thermodynamic and
geometrothermodynamic properties of the Bardeen regular black hole solution.
First, we present a brief review of the GTD formalism, emphasizing only the final results and the importance of considering the quasi-homogeneous character of black holes. 
Then, we used the fundamental equation of the Bardeen black hole and the first law of black hole thermodynamics to derive all the relevant variables such as the temperature and the magnetic potential dual to the magnetic charge. The temperature turns out to be physical in a range that limits the entropy values in terms of the magnetic charge.

To simplify the geometrothermodynamic analysis, we use the mass as the thermodynamic potential and derive the explicit form of the three families of Legendre invariant metrics previously derived in GTD. It turns out that the three metrics have non-zero curvature, indicating the presence of thermodynamic interaction. To interpret the geometric properties of the equilibrium space of the Bardeen black hole, we compare the results obtained in GTD with those derived from classical black hole thermodynamics. We conclude that  
the metric $g^I_{ab}$ is characterized by curvature singularities that are interpreted as due to a violation of the stability conditions of the black hole when considered as a thermodynamic system. 
The second metric $g^{II}_{ab}$ leads to curvature singularities that are interpreted as second-order phase transitions by comparing them with the divergences of the heat capacity. The metric $g^{III}$ is regular for all physically allowed values for the entropy and magnetic charge.
We conclude that the presence of the magnetic charge in the Bardeen black hole drastically changes the stability and phase properties of the system because it introduces a new physical interval in which the black hole can become stable and undergo phase transitions. 

In addition, we performed a detailed analysis of the critical exponents of the heat capacity and the GTD curvature scalars of the Bardeen black hole. We found 
that the heat capacity diverges as $ |C_q|
\sim |T-T_c|^{-1/2}$ 
and the curvature of the metric $g^{II}$ as i.e., $ R^{II} 
\sim |T-T_c|^{-1}$, so that the scaling relation $R^{II}\propto C_q^{\,2}$ is satisfied. In this way, we have shown that the thermodynamic curvature contains information and about the location of the Davies phase transition and its critical behavior.
Our results also show that GTD is a geometric formalism that can be used to derive the physical properties of thermodynamic systems in an invariant way and leads to results that are compatible with classical thermodynamics. 

The present analysis has been restricted to the Bardeen regular black hole, which provides a representative example to investigate the role of GTD in the description of critical phenomena. An interesting direction for future research is to extend the present approach to other families of regular black holes and to determine whether the scaling behavior of the thermodynamic curvature and the associated critical exponents represent universal features of Legendre invariant thermodynamic geometry or depend on the particular black-hole solution under consideration.

\section*{Acknowledgements}
The work of MNQ was carried out within the scope of the project CIAS 3918 supported by the Vicerrectoría de Investigaciones de la Universidad Militar Nueva Granada - Vigencia 2024. This work was partially supported by DGAPA-PAPIIT UNAM, Grant No. 108225, and 
Conahcyt, grant No. CBF-2025-I-243

\end{document}